\documentclass[12pt,a4paper]{article}
\usepackage{epsfig}
\usepackage{amssymb}
\newcommand{\vect}[1]{{\bf #1}}
\newcommand\sfrac[2]{{\textstyle{\frac{#1}{#2}}}}
\newcommand\deriv[2]{\displaystyle\frac{\partial #1}{\partial #2} }
\newcommand\col[2]{ \left(\begin{array}{c}#1\\#2\end{array}\right) }

\newcommand{\la}{\;
  \raise0.3ex\hbox{$<$\kern-0.75em\raise-1.1ex\hbox{$\sim$
  }}\;\hskip-2pt }
\newcommand{\ga}{\;
  \raise0.3ex\hbox{$>$\kern-0.75em\raise-1.1ex\hbox{$\sim$
  }}\;\hskip-2pt }

\newcommand{\slashint}{\setminus\!\!\!\!\!\!\int}
\newcommand{\cm}{\,{\rm cm}}

\newcommand{\kpc}{\,{\rm kpc}}

\newcommand{\p}{\,{\rm pc}}

\newcommand{\s}{\,{\rm s}}
\newcommand{\yr}{\,{\rm yr}}

\newcommand{\beq}{\begin{equation}}
\newcommand{\eeq}{\end{equation}}
\newcommand{\beqarr}{\begin{eqnarray}}
\newcommand{\eeqarr}{\end{eqnarray}}
\newcommand{\barr}{\begin{array}}
\newcommand{\earr}{\end{array}}
\newcommand{\bcent}{\begin{center}}
\newcommand{\ecent}{\end{center}}

\newcommand{\pder}[3]{\frac{\partial^{#3}#1}{\partial #2^{#3}}}
\renewcommand{\vec}[1]{{\bf #1}}

\def\today{
\number\day\space
\ifcase\month\or
January\or February\or March\or April\or May\or June\or
July\or August\or September\or October\or November\or December\fi
\space\number\year}

\date{}

\title{
   \Large{\bf NON-LOCAL EFFECTS IN THE MEAN-FIELD DISC DYNAMO.\\
   II.  NUMERICAL AND ASYMPTOTIC SOLUTIONS}
}

\author{
   ASHLEY P.~WILLIS\footnote{
   Now at School of Earth Sciences, University of Leeds.}
    and ANVAR SHUKUROV\\
   {\normalsize\it School of Mathematics and Statistics,
        University of Newcastle,}\\
   {\normalsize\it Newcastle upon Tyne, NE1~7RU, UK}\\[10pt]
   ANDREW M.~SOWARD\\
   {\normalsize\it Department of Mathematical Sciences, University of Exeter,}\\
   {\normalsize\it North Park Road, Exeter, EX4 4QE, UK}\\[10pt]
   DMITRY SOKOLOFF\\
   {\normalsize\it Department of Physics, Moscow University, Moscow
   119992, Russia}\\[10pt]
   \bigskip
   \centerline{
      {\normalsize\it (Received \today; in final form \hspace{2cm})}
   }
}

\begin{document}

\maketitle

\begin{abstract}
The thin-disc global asymptotics are discussed for
axisymmetric mean-field dynamos with vacuum boundary
conditions allowing for non-local terms arising from a finite
radial component of the mean magnetic field at the disc
surface.  This leads to an integro-differential operator in
the equation for the radial distribution of the mean magnetic field
strength, $Q(r)$ in the disc plane at a distance $r$ from its centre;
an asymptotic form of its solution
at large distances from the dynamo active region is obtained.
Numerical
solutions of the integro-differential equation confirm that the non-local
effects act similarly to an enhanced magnetic diffusion.  This leads to
a wider
radial distribution of the eigensolution and faster propagation of magnetic
fronts, compared to solutions with the radial surface field neglected.
Another result of non-local effects is a slowly decaying algebraic tail of
the eigenfunctions outside the dynamo active region, $Q(r)\sim r^{-4}$, which
is shown to persist in nonlinear solutions where $\alpha$-quenching is
included.  The non-local nature of the solutions can affect the radial profile
of the regular magnetic field in spiral galaxies and accretion discs
at large distances from the centre.

\bigskip
\noindent
KEY WORDS: Mean-field dynamos, thin-disc asymptotics, boundary
conditions, galactic magnetic fields, accretion discs.

\end{abstract}

\section{Introduction}

Thin-disc asymptotics, applicable to spiral galaxies and accretion
discs, have been a useful tool in the studies of the origin of
large-scale magnetic fields in these objects. The small parameter
naturally arising in a thin disc is its aspect ratio
\begin{equation}
\lambda\,\equiv\,h_0/R_0\,\ll\,1\,,
\label{smalllambda}
\end{equation}
where $h_0$ and $R_0$ are the characteristic half-thickness and radius of the
disc respectively. The height of the disc surface and the radius of its edge,
understood as positions where suitable boundary conditions for the disc's
magnetic field can be reasonably applied, are $h=O(h_0)$ and $R=O(R_0)$.
In a domain that has no sharp boundaries (e.g., a plasma layer in hydrostatic
equilibrium along the vertical direction and centrifugal equilibrium along
radius), $h_0$ and $R_0$ can be identified with the scale height and the
radial scale length, respectively.

We adopt cylindrical polar coordinates $(r,\phi,z)$, in which
$z$ measures distance parallel to the rotation axis
and assume that our system is axisymmetric, independent of the
azimuthal angle $\phi$. We assume that the mean motion is
dominated by differential rotation and is approximated by
the velocity $\vec{V}=(0,V,0)$.
We consider an axisymmetric magnetic field whose poloidal components
can be expressed in terms of the azimuthal component $A$
of the vector potential:
\begin{equation}
 \vect{B}=\left(-\,\frac{\partial A}{\partial  z},\,\,B,\,\frac{1}{r}\,\deriv{}{r}(rA)\right).
\label{BA}
\end{equation}
The magnetic field satisfies the mean-field dynamo equation
\begin{equation}
\deriv{\vec{B}}{t}
=\nabla\times\left(\vec{V\times B}+\alpha\vec{B}
-\beta\nabla\times\vec{B}\right)
\label{MFD}
\end{equation}
(see Moffatt, 1978; Krause and R\"adler, 1980),
in which $\alpha$ and $\beta$ are turbulent transport coefficients
responsible for the $\alpha$-effect and turbulent magnetic diffusion,
respectively. We will consider the linear problem with
$V$,  $\alpha$ and $\beta$ assumed given and independent
of time $t$, for which solutions may be sought proportional
to $\exp(\Gamma t)$, where the constant $\Gamma$ is the growth rate.

The dynamo equation (\ref{MFD}) has been solved with a non-local boundary 
condition, possible for an infinite slab, using the Bullard-Gellman 
(1954) formalism (Raedler and Wiedemann, 1990).  Traditionally, 
however, the boundary conditions used in conjuction with
(\ref{MFD}) are the so-called vacuum boundary
conditions. If there are no electric currents outside the disc, then
$\nabla\times\vect{B}=\bf0$ there, and the magnetic field may be expressed
as the gradient of a potential in the form $\vect{B}=-\nabla\Phi$.
Thus the axial symmetry implies that the azimuthal magnetic field vanishes,
\begin{equation}    \label{Bbc}
B=0\qquad\mbox{outside the disc}
\end{equation}
(see, e.g., Zeldovich {\it et al.,} 1983, p.~151),
while the remaining meridional magnetic field is potential $\nabla^2\Phi=0$
since $\nabla\cdot\vect{B}=0$.
Our solution of the potential problem relies on two approximations.
Firstly, we assume that the disc is so thin that its
surface is essentially at $z=0$.
In this way we may obtain, using the Green's function, 
a solution of the Neumann problem for $\Phi(r,z)$ in the
upper half plane $z>0$ in  which the value of
$B_z(r,0_+)=-\partial\Phi/\partial z(r,0_+)$ is specified.
This Green's function solution may by used to calculate
$B_r(r,0_+) =-\partial\Phi/\partial r(r,0_+)$. In terms of the vector potential $A$
for the radial and axial magnetic field components
$B_r=-\partial A/\partial z$ and $B_z=(1/r)\partial(rA)/\partial r$,
the resulting expression for $B_r$ in terms of $B_z$, both at $z=0_+$, may be written
in the form
\begin{equation}
\deriv{A}{z}(r,0_+) - {\cal L}\left\{A(r,0_+)\right\} = 0\,,
\label{Phibc}
\end{equation}
where the operator
\begin{equation}
{\cal L}\left\{A(r,0_+)\right\}=
\deriv{}{r}\left[\int_0^\infty G(r,r')\,\deriv{}{r'}\Bigl(r'A(r',0_+)\Bigr)\,dr'\right]
\label{intopPhi}
\end{equation}
involves the Green's function
\begin{equation}
 G(r,r') \equiv \frac{1}{2\pi}\int_{-\pi}^{\pi}
\frac{d\theta}{\sqrt{r^2-2rr'\cos\theta+r'^2}}\,.
\label{Greentheta}
\end{equation}
Secondly, we assume that the magnetic field is localised above the disc and
is negligible beyond $r=R$. That means that the upper limit $\infty$ in the
integral (\ref{intopPhi}) is interchangeable with $R$. Then (\ref{Phibc}) is to be regarded
as an equation that determines $B_r$  as a nonlocal integral function
of $B_z$ on the disc surface $z=0_+$ for vacuum boundary conditions.
Indeed for the quadrupolar magnetic fields, which concern us
in this paper, we also have $B_z(r,0)=0$ for $r>R$. For that case, no
approximations are involved in interchanging $\infty$ with $R$.

In a previous paper, Priklonsky {\it et al.} (2000, hereafter referred
to as Paper I) adopted the slightly different formulation
\beq
{\cal L}\left\{A(r,0_+)\right\}=\,\frac{1}{r}\,
\int_0^\infty W(r,r')\,\,
      \deriv{}{r'}\left[\frac{1}{r'}\,\deriv{}{r'}\Bigl(r'A(r',0_+)\Bigr)\right] dr'
\label{intopA}
\end{equation}
of the integral operator based on the solution of
$(\nabla^2 -r^{-2})A=0$ in the vacuum region outside the disc, where 
$W(r,r')$ will be given below, see (\ref{GreenJ}).
It is readily established by integration by parts
that the operators (\ref{intopPhi}) and (\ref{intopA}) are equivalent
provided that the two kernels $G(r,r')$ and $W(r,r')$ are related by
\begin{equation}
r\,\deriv{}{r}[G(r,r')]\,=\,-\,\frac{1}{r'}\,\deriv{}{r'}[W(r,r')]\,.
\label{GWrel}
\end{equation}
It is the latter formulation (\ref{intopA}) that we employ to obtain our
numerical results.
The numerical implementation involves a regularisation procedure to
allow for the singularity of the kernel,
$W\approx -\pi^{-1}\sqrt{rr'}\ln |r-r'|$ at $r=r'$.

Inside thin discs, $|z|<h(r)$ on $0\le r<R$, where $h$ may be
a function of $r$, WKB type solutions of (\ref{MFD}) can be constructed in the form
\begin{equation}
\col{B}{A}=\exp{(\Gamma t)}\left[Q(r)\col{b(z;r)}{a(z;r)}+\ldots\right]
\qquad\quad\mbox{with }\qquad
Q(r)\,=\,{\widehat Q}(r/\lambda^{s})
\label{asymptexp}
\end{equation}
(Soward, 1978, 1992a,b; Ruzmaikin {\it et al}., 1985, 1988).
The local $z$-structure on the short length scale $h$ of the disc thickness,
is determined by $(b,a)$ normalised at our convenience; here $z$ is measured
in units of $h$ so that the boundary is located at $z=\pm 1$. The
corresponding time unit used to measure the inverse growth rate $1/\Gamma$ is
$h_0^2/\beta$. The modulation on the long radial length scale $R_0$, adopted
as our unit of $r$, is quantified by the amplitude $Q(r)$. In our thin disc
($\lambda\ll 1$) the solution ${\widehat Q}(r/\lambda^{s})$ is, in fact, usually modulated on a radial scale
$\lambda^sR_0$ intermediate between $R_0$ and $h_0$, where $s$ is a constant
satisfying $0<s<1$. The scale $\lambda^s R_0$ is large compared to the disc
width, $h=O(\lambda R_0)$, but small compared to the disc radius,
$R$, a quantity of order $R_0$.  The actual value taken by $s$ depends on the nature
of the boundary conditions. It is $s=1/3$ for our non-local boundary condition
(\ref{Phibc}) but may take other values if different boundary conditions are
adopted; e.g., the value $s=1/2$ found by Ruzmaikin {\it et al.} (1988) for a
`local' boundary condition as we explain below in relation to (\ref{local}).

We note that generally the growth rate $\Gamma$ of the dominant modes in a thin
disc is real, when  $\pm\alpha>0$ for $\pm z>0$.
Raedler and Wiedemann (1990) considered non-local boundary conditions 
for an infinite slab, also finding that non-oscillatory modes are the 
most easily excited. 
The asymptotic expansion of the form (\ref{asymptexp})
applies to modes of quadrupolar parity
that dominate in galactic discs; dipolar modes,
perhaps prevailing in accretion discs (e.g., Moss and Shukurov, 2003),
have been discussed by Soward (1978, 1992a,b).

The lowest-order approximation in $\lambda$ yields a
boundary value problem for $(b,a)$
that only involves derivatives in $z$ and so is local in $r$.
The local problem yields the local eigenfunction $(b,a)$ together
with its eigenvalue, namely the `local growth rate' $\gamma(r)$
(see Paper I for details).
Relative to local coordinates, in which the units of length for
$z$ and $r$ are $h$ and $R_0$ respectively  ($z\to z/h$ and $r\to r/R_0$),
the local eigenfunction determines a linear relation
between $\partial a/\partial z(1;r)$ and $a(1;r)$
at the disc surface.
We use dimensionless coordinates $z\to z/h$ and $r\to r/R_0$, 
keeping the notation $z$ and $r$ for them, unless stated otherwise.
Used in conjunction with
the boundary condition (\ref{Phibc}), the linear relation
leads at next order to an amplitude equation governing $Q(r)$.
Written in the form derived in Paper I, it is
\beq
   [\Gamma \,-\,\gamma(r)]\,q(r)\,=\,\lambda\eta(r)\,{\cal L}\left\{q(r)\right\},
\label{nonlocal}
\eeq
where the factor $\lambda$ reflects the fact that the radial unit
of distance $R_0$ is a factor $1/\lambda$ larger than the axial unit
$h$,
the integral operator $\cal L$ is defined by (\ref{intopA}) with
kernel $W(r,r')$ or (\ref{intopPhi}) with kernel $G(r,r')$ and
\beq
    q(r)=Q(r)a(1;r)\;,
\qquad
    \eta(r)=\frac{a(1,r)a_*(1,r)}{\langle\vect{X},\vect{X}_*\rangle}\;,
\qquad
    \vect{X}=\left[\begin{array}{l}b(z;r)\\a(z;r)\end{array}\right].
\label{qeta}
\eeq
Here
$\vect{X}$ is the eigenvector of the lowest-order boundary value problem,
the asterisk denotes the eigenvector of its adjoint problem, while
\[
\langle\vect{X},\vect{X_*}\rangle=\int_0^1 \vect{X}\cdot\vect{X_*}\,dz\;.
\]
The solution of (\ref{nonlocal}) subject to the boundary conditions
\beq
q(0)=0\,\qquad\mbox{and} \qquad q\to 0\quad\mbox{as}\quad r\uparrow\infty
\label{bcsq}
\eeq
provides yet another eigenvalue problem, for which the eigenvalue is the
 growth rate $\Gamma$ and the eigenfunction is $q(r)$
which determines the radial $r$-modulation $Q$.

Soward (1978, 1992a,b) developed a systematic asymptotic theory for
the marginally stable dynamo eigensolutions in a
thin slab that includes the non-local coupling between its distant
parts via the surrounding vacuum, as described by the integral operator in
(\ref{nonlocal}) but in simplified
local Cartesian rather than axisymmetric form.
Though this theory was originally formulated in terms of
Fourier transforms, real space formulation is given in
(\ref{hatgammaex}) as obtained in Soward (2003). He pointed out, in particular, that
the non-local coupling produces slowly decaying (algebraic) tails in the
magnetic field radial distribution far away from the localization region of an
eigenfunction. The corresponding theory with a full account for the
cylindrical geometry and with allowance for growing (rather than steady) solutions
was proposed in Paper I.

The integral term in (\ref{nonlocal}) was neglected by Ruzmaikin {\it et al.} (1988),
who were interested mainly in solutions
within the dynamo active region, where the non-local external potential coupling
may not be very important, so the integral term in (\ref{Phibc}) was neglected.
Instead these authors included local internal coupling by diffusion,
which leads to the radial dynamo equation
\begin{equation}
   [\Gamma -\gamma(r)] Q = \,\lambda^2
\pder{}{r}{}\left[\frac{1}{r}\pder{}{r}{}\Bigl(rQ(r)\Bigr)\right]
\label{local}
\end{equation}
similar to (\ref{nonlocal}), but with the integral term replaced by
 the diffusion operator.
The value of the exponent $s$ in (\ref{asymptexp}) is
sensitive to this difference. If (\ref{local}) is used, then
$s=1/2$ (Ruzmaikin {\it et al.}, 1985, 1988), whereas non-local
asymptotics have $s=1/3$ (Soward, 1978).

In this paper we discuss numerical solutions of the non-local radial
equation (\ref{nonlocal}), both in the kinematic regime and in a
nonlinear regime with heuristically chosen nonlinearity. We compare
solutions of the non-local equation (\ref{nonlocal})
with those of the local equation (\ref{local}), where the integral term
is replaced by the diffusion operator, and confirm that the non-local
coupling produces effects similar to enhanced radial diffusion. We
also demonstrate that the algebraic tail of magnetic field outside
the dynamo active region persists in nonlinear solutions.

Before continuing it must be emphasised that though it was convenient in
(\ref{qeta}) to measure the $z$-distance in units of $h$ for the internal
disc magnetic field, this is quite inappropiate in the vacuum region outside.
So for the remainder of our paper both $z$ and $r$ will be measured in units of $R_0$, 
though we continue to write $a(1,r)$ for the value of $a$ on the disc surface.

\section{The Integral Kernel}            \label{sect:kernel}

In this section we develop various forms for the equivalent integral operators
(\ref{intopPhi}) and (\ref{intopA}) employed in (\ref{nonlocal})
which may be utilised effectively in our numerical methods
as well as aid our understanding of the asymptotic results.

\subsection{Properties of $G(r,r')$ and $W(r,r')$}

We begin by noting that the kernels $G(r,r')$ and $W(r,r')$
in (\ref{intopPhi}) and (\ref{intopA}) have the integral representations
\begin{equation}
 G(r,r')\,=\,\int_0^\infty J_0(kr)\,J_0(kr')\,dk\,,
\quad\qquad
 W(r,r')\,=\,rr'\,\int_0^\infty J_1(kr)\,J_1(kr')\,dk
\label{GreenJ}
\end{equation}
in terms of the Bessel functions $J_0(x)$ and $J_1(x)$ of the first kind.
The equivalence of the former for $G(r,r')$ with (\ref{Greentheta})
follows from Gradshteyn and Ryzhik's (2000), (3.674.1) and (6.512.1), and
Abramowitz and Stegun's (1965), (17.3.9); henceforth referred to as
G\&R and A\&S respectively.
Further it is readily established using the elementary properties of
the Bessel functions and their derivatives
that the two kernels $G(r,r')$ and $W(r,r')$
are related by (\ref{GWrel}).

Significantly the kernels have the alternative representations
\begin{equation}
\left.
\begin{array}{rcl}
G(r,r')&=&\displaystyle{\frac{2}{\pi r}\,K(m)}\,,  \\ [1em]
W(r,r')&=&\displaystyle{\frac{2r}{\pi}\,m\left[K(m)+2(m-1)\deriv{K}{m}(m)\right]}
\end{array}
\right\}
\qquad \mbox{for}\quad r'<r
\label{GWEI}
\end{equation}
in terms of the complete elliptic integral of the first kind
\begin{equation}
   K(m) = \int_0^{\pi/2} \frac{d\theta}{\sqrt{1-m\sin^2\theta}}
\qquad\mbox{and}\qquad
m\,=\,\left(\frac{r'}{r}\right)^2\,
\label{Kdef}
\end{equation}
(see G\&R (6.512.1), A\&S (15.2.7), (17.3.9)).
Direct verification that $G(r,r')$ and $W(r,r')$ defined by (\ref{GWEI})
satisfy the relations (\ref{GWrel}) follows from the identity
$4(d/dm)[m(1-m)dK/dm]=K$ (see G\&R (8.124.1)).
The forms for $r'>r$ are obtained using the symmetry relations
\begin{equation}
G(r,r')\,=\,G(r',r)\,,\qquad\qquad W(r,r')\,=\,W(r',r)\,.
\label{Symm}
\end{equation}

The asymptotic representations of the kernels are readily determined by
documented properties of the complete elliptic integral. In particular
use of the formula $2K(m)\approx\ln[16/(1-m)]$ valid when $1-m\ll 1$
(see A\&S (17.3.26))
shows that the kernels  $G(r,r')$ and  $W(r,r')$ have the expansions
\begin{equation}
\left.
\begin{array}{rcl}
G(r,r')&\approx&
\displaystyle{\frac{1}{\pi r}\ln\left(\frac{8r}{r-r'}\right)}, \\[1em]
W(r,r')&\approx&
\displaystyle{\frac{r}{\pi}\left[\ln\left(\frac{8r}{r-r'}\right)-\,1\right]}
\end{array}
\right\}
\qquad \mbox{for}\quad 0<r-r'\ll r\,.
\label{Wnear}
\end{equation}
Likewise use of the formula $K(m)\approx (\pi/2)(1+m/4)$ valid when $m\ll 1$
(see A\&S (17.3.11))
shows that the kernel $G(r,r')$ has the expansion
\begin{equation}
G(r,r')\,\approx\,
\displaystyle{\frac{1}{r}\,+\,\frac{r'^2}{4r^3}}
\qquad \mbox{for}\quad r'\ll r\,.
\label{Gfar}
\end{equation}
Together these results determine the important asymptotic forms
\begin{equation}
\frac{\partial^2 G}{\partial r \partial r'}(r,r')\,\approx
\left\{
\begin{array}{ll}
\displaystyle{-\,\frac{1}{\pi r(r-r')^2}}\qquad &\mbox{for} \quad 0<r-r'\ll r\,,  \\ [1em]
\displaystyle{-\,\frac{3r'}{2r^4}}\qquad &\mbox{for} \quad  r'\ll r\,.
\end{array}
\right.
\label{GnWf}
\end{equation}

\subsection{Asymptotic properties of localised solutions}
In this section we discuss asymptotic forms of the solutions far from
the dynamo active region. In particular, we clarify the nature of the slow,
algebraic (rather than exponential) decay of the solution with cylindrical radius
at large radii.

When $q(r)$ is localised about some point $r_m$ on a length scale
$\epsilon$, where $\epsilon\ll r_m$, we integrate (\ref{intopPhi}) by parts
to obtain [noting that $A(r,1)=q(r)\exp{(\Gamma t)}$]
\beq
{\cal L}\left\{q(r)\right\}=\,-\,
\deriv{}{r}\left[\,\,\slashint_0^\infty
  r'q(r')\,\deriv{}{r'}[G(r,r')]\,dr'\right] \qquad\qquad\quad
\label{intopnfA}
\eeq
\begin{equation}\quad\qquad\qquad
\approx\,-\,{\cal Q}\,r_m\frac{\partial^2 G}{\partial r \partial r_m}(r,r_m)
\qquad \mbox{for}\quad |r-r_m|\gg \epsilon\,,
\label{intopnf}
\end{equation}
where the slashed integral sign indicates principal part
[needed here because $\partial G/\partial r'\sim (r-r')^{-1}$] and
\begin{equation}
{\cal Q}\,\equiv\,\int_0^\infty q(r)\,dr\,.
\label{QAdef}
\end{equation}
This  result may be used in conjunction with (\ref{nonlocal})
to obtain the far field behaviours
\begin{equation}
q(r)\,\approx\,-\,\frac{\lambda \eta(r){\cal Q}}{\Gamma-\gamma(r)}
\,r_m\frac{\partial^2 G}{\partial r \partial r_m}(r,r_m)
\qquad \mbox{for} \quad |r-r_m|\gg \epsilon
\label{gex}
\end{equation}
of the modulation amplitude and the radial magnetic field
\beq
B_r(r,1)\equiv-\frac{1}{\lambda}\deriv{A}{z}(r,1)\,\approx
\exp(\Gamma t){\cal Q}
\,r_m\frac{\partial^2 G}{\partial r \partial r_m}(r,r_m)
\qquad \mbox{for}\quad |r-r_m|\gg \epsilon
\label{intopnfB}
\eeq
on the disc surface, as defined by (\ref{Phibc}).
Furthermore from (\ref{GnWf}) we may employ the limiting forms
\begin{equation}
r_m\frac{\partial^2 G}{\partial r \partial r_m}(r,r_m)
\,\approx
\left\{
\begin{array}{ll}
\displaystyle{\,\frac{1}{\pi(r-r_m)^2}}\qquad &\mbox{for} \quad r_m\gg |r-r_m|\gg \epsilon\,,  \\ [1em]
\displaystyle{\,\frac{3r_m^2}{2r^4}}\qquad &\mbox{for} \quad  r\gg r_m\,.
\end{array}
\right.
\label{diquadru}
\end{equation}
The implication of the $r$-dependencies in (\ref{intopnfB}) and (\ref{diquadru}) is
that the response of the magnetic field in vacuum outside the disc corresponds to
a line dipole source when $r_m\gg |r-r_m|\gg \epsilon$ but corresponds to
a quadrupole source when $r\gg r_m$.

To illustrate the idea we consider the `local growth rate'
\begin{equation}
\gamma(r)=\gamma_m - \sfrac{1}{2}\,|\gamma_m''|\,(r_m-r)^2
\qquad \hbox{on}\quad 0\le r<\infty\;,
\label{gammaex1}
\end{equation}
so that the influence of $\gamma(r)$ is dominated by its behaviour
near $r_m$,
where $\gamma_m$ and $\gamma_m''$ are constants.
Then the amplitude $q(r)$ governed by (\ref{nonlocal}) satisfies
\begin{equation}
\left[(\Gamma-\gamma_m) + \sfrac{1}{2}\,|\gamma_m''|\,(r_m-r)^2\right]q(r)
\,=\,\lambda\eta_m{\cal L}\{q(r)\}\,,
\label{gammaex2}
\end{equation}
where $\eta_m\equiv \eta(r_m)$. Upon setting
\begin{equation}
r\,=\,r_m+\epsilon \varpi\,,
\qquad
\Gamma\,=\,\gamma_m + \sfrac{1}{2}\epsilon^2\,|\gamma_m''|\,\delta
\qquad
\mbox{with}\quad
\epsilon\,=\,(2\lambda\eta_m/|\gamma_m''|)^{1/3},
\label{scaling}
\end{equation}
the integral equation  (\ref{gammaex2}) may be approximated for
$\varpi=O(1)$ with the help of (\ref{Wnear}) and (\ref{intopnfA})
to yield
\begin{equation}
\left(\,\delta + \varpi^2\,\right){\widehat q}(\varpi)
\,\approx\,\frac{1}{\pi}\,
\slashint_{-\infty}^{\infty}
\frac{1}{\varpi'-\varpi}\,\frac{d{\widehat q}}{d\varpi'}(\varpi')\,d\varpi'\,,
\label{hatgammaex}
\end{equation}
where $\widehat{q}(\varpi)=q(r)$. Here, since $\epsilon\ll1$,
the integration over $0\leq r<\infty$ is equivalent asymptotically
to $-\infty<\varpi<\infty$ on the basis that
$\widehat{q}\to 0$ as $|\varpi|\equiv |r-r_m|/\epsilon\to\infty$.

The lowest eigenvalue $\delta$ of the eigenvalue problem
(\ref{hatgammaex}) given by Soward (2003) is
$\delta\approx -1.0188$, namely the first zero of the derivative of Airy's
function, $\mathrm{Ai}'(\delta)=0$. For that value we have the eigensolution
\beq
{\widehat q}(\varpi)\,=\,\frac{\widehat{\cal Q}}{\pi\mathrm{Ai}(\delta)} \int_0^\infty \mathrm{Ai}(k+\delta)\cos k\varpi\,dk\;,
\label{intsol}
\eeq
where $\mathrm{Ai}(\delta)\approx 0.53565$ and
\beq
{\widehat{\cal Q}}\,=\,\int_{-\infty}^{\infty}{\widehat q}(\varpi)\,d\varpi\,.
\label{calQ}
\eeq
The large $\varpi$ behaviour of ${\widehat q}$ is
\beq
{\widehat q}\,\approx\,\frac{\widehat{\cal Q}}{\varpi^4}\,.
\label{largevarpi}
\eeq
This result is consistent with (\ref{gex}) and (\ref{diquadru}) provided that
$r_m\gg |r-r_m|\gg \epsilon$.

\section{Numerical results}
In this section we discuss numerical solutions of the non-local radial
equation (\ref{nonlocal}) and assess the importance and implications of the
non-local effects by comparing the solutions with those of its local
`approximation', (\ref{local}). Our numerical solution of
(\ref{nonlocal}) employs the integral kernel $W(r,r')$ defined by
(\ref{GWEI}). To that end $K(m)$ can be computed iteratively using the
procedure of the arithmetic-geometric mean described in Sect.~17.6 of A\&S,
which displays phenomenal convergence. We also need to compute $dK/dm$; for
this, we use a modification of the above procedure described in Appendix~A.
Numerical solutions of the linear equation (\ref{nonlocal}) and its nonlinear
version, which we introduce in Section \ref{SS}, involve a further
regularization procedure described in Appendix~B to contend with the
singularity of the kernel.

The local dynamo problem at a fixed $r$, that yields the local growth rate
$\gamma(r)$ and the local eigenfunction $\left(b(z;r),a(z;r)\right)$,
will not be solved here.
Instead a suitable local growth rate, $\gamma(r)$, is prescribed
and we take
\beq
a(1;r)=1\,,\qquad
 \eta(r)=\widetilde\eta(r)=1\,,
\qquad
\mbox{so that} \quad
q(r)=Q(r)\,.
\label{choice}
\eeq
We adopt the boundary conditions
\beq
Q(0)\,=\,Q(R)\,=\,0\,,
\label{bcQ}
\eeq
where $R$ is the disc radius.
Since our nonlocal integral equation (\ref{nonlocal})
is derived for a disc of infinite radius, this may lead to
an artificial boundary layer at $r=R$ if $R$ is not large enough.
Thus it is important that our magnetic
field is localised and negligible for $r\ge R$.
In the following subsections we consider examples of the cases
in which $\gamma(r)$ is maximised on the symmetry axis $r_m=0$
and also at a finite distance $r_m>0$ from it.

\subsection{Linear solutions for the case $r_m=0$}      \label{LS}

\begin{figure}
   \bcent
         \epsfig{figure=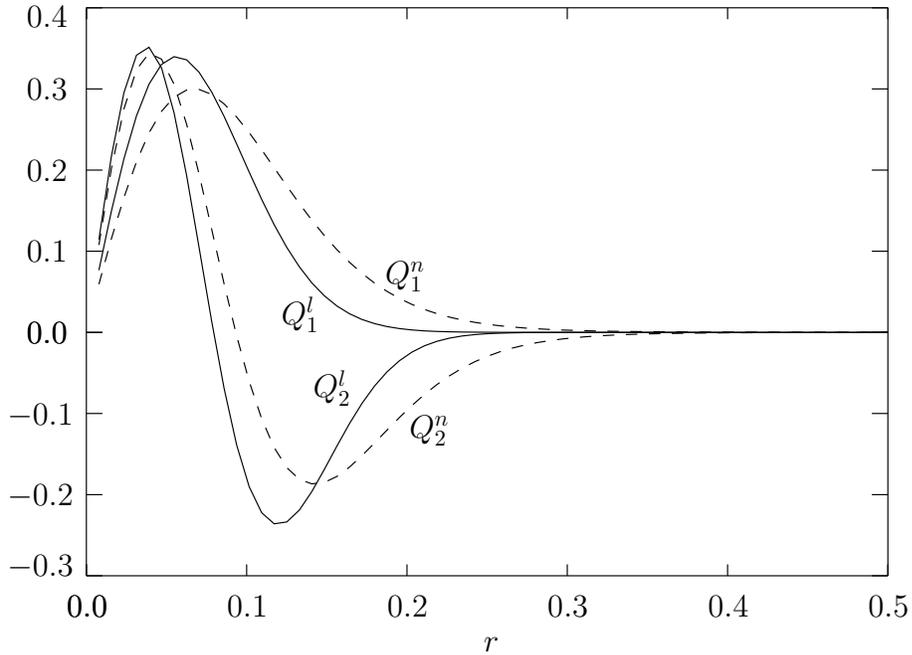, scale=0.85}
      \caption{ \label{fig:emodes}
The first two eigenfunctions $Q_1$ and $Q_2$ of the local and non-local
dynamo equations (\protect\ref{local}) and
(\protect\ref{nonlocal}) respectively, for the case
$\lambda=0.01$ and $\gamma(r)$ of the form (\ref{parab}) with 
$\gamma_\mathrm{m}=10$,
$r_0=1$ and $R=2$.
The local (non-local) eigenfunctions $Q$ are shown with solid (dashed)
curves and labelled with the superscript $l$ ($n$).
      }
   \ecent
\end{figure}

\begin{figure}
   \bcent
      \begin{tabular}{c}
         \vspace{-1cm}
         \epsfig{figure=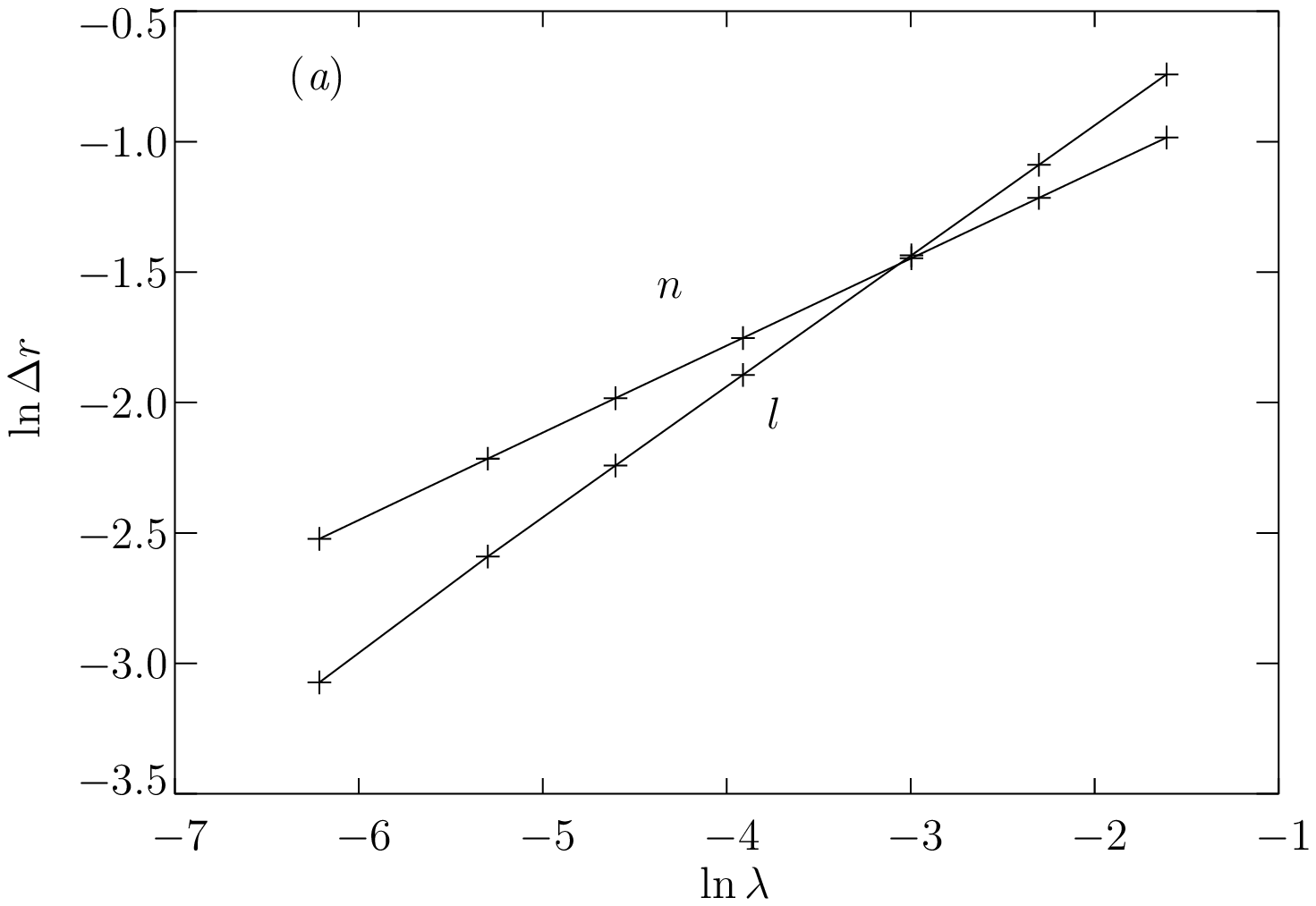, scale=0.80}
      \\[30pt]
         \epsfig{figure=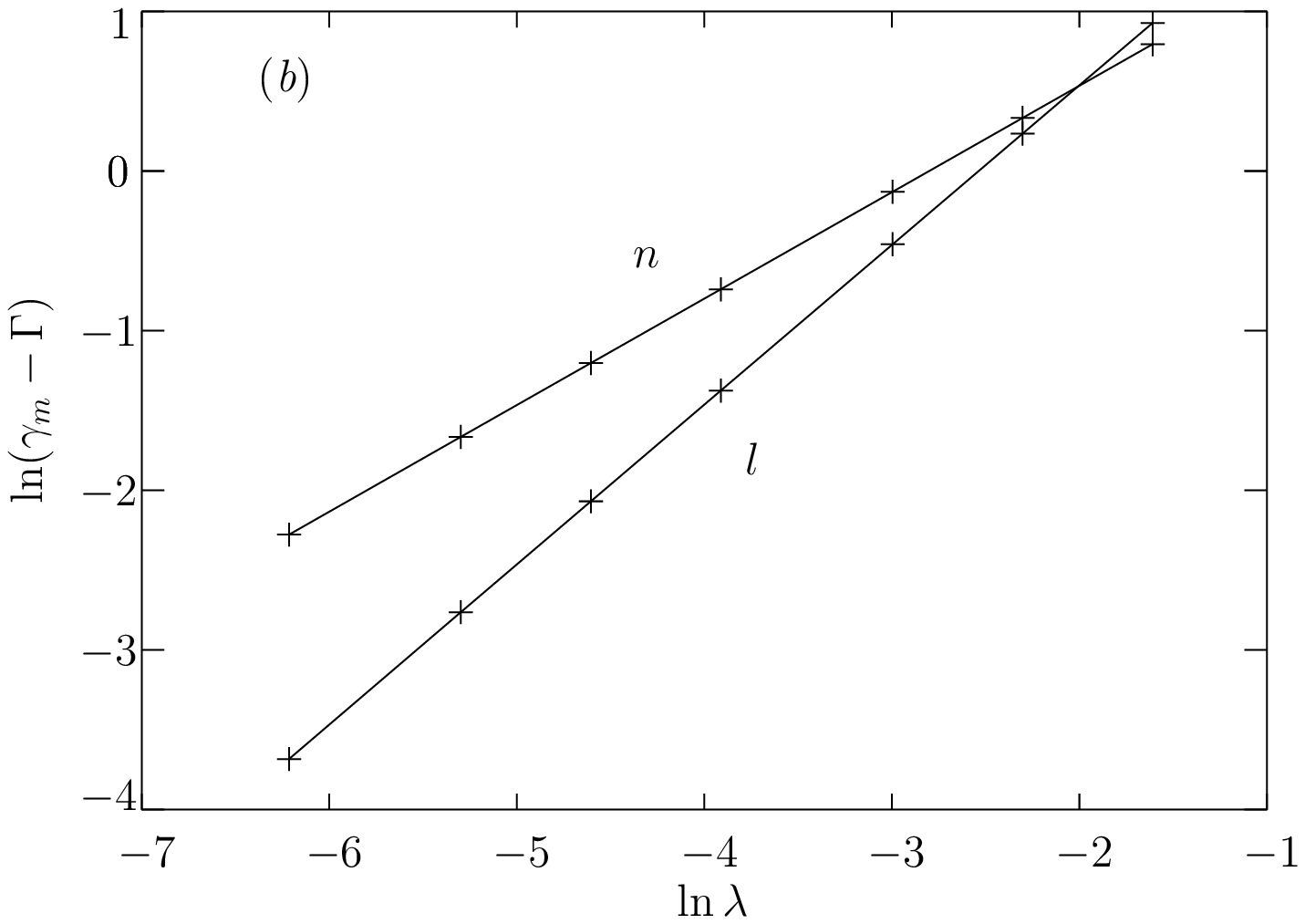, scale=0.80}
      \end{tabular}
   \ecent
   \caption{ \label{fig:lambda-var}
The dependencies of the leading eigensolutions on the dimensionless
disc thickness $\lambda$ showing the variations on a log-log scale:
{\bf(a)} the width of the eigenfunctions $\Delta r$ (i.e., the distance between positions
where $Q(r)$ decreases to $e^{-1}$ of its maximum value)
and {\bf(b)} the decrement $\gamma_{m}-\Gamma$ of the eigenvalues.
The local and non-local solutions are distinguished by labels
{\it l\/} and {\it n}, respectively.
   }
\end{figure}

The two leading eigenfunctions for the nonlocal (\ref{nonlocal})
and local (\ref{local}) eigenvalue  problems are shown in
Fig.~\ref{fig:emodes} for the parabolic profile
\beq                    \label{parab}
   \gamma(r) =
   \left\{
      \barr{ll}
         \gamma_\mathrm{m}(1-r^2/r_0^2)\;, & \quad r\le r_0\;, \\
         0\;,         & \quad r>r_0
      \earr
   \right.
\eeq
with $\gamma_m=10$, $r_m=0$, $r_0=1$
and $R=2$. The non-local eigenfunctions are broadened versions
of the local solutions but their qualitative similarity is evident
in Fig.~\ref{fig:emodes}.
Such broadening of the eigenfunctions may be achieved from
an anomalous, enhanced diffusion.
For $\lambda$ in the range applicable to spiral galaxies, namely
$0.01\la \lambda \la 0.1$, the eigenvalues $\Gamma$ (growth rate)
resulting from the non-local problem are smaller,
while the eigenfunctions are generally wider
-- see Fig.~\ref{fig:lambda-var}.
As might be expected, the non-local effects become weaker as $\lambda$
decreases, i.e., as the disc becomes thinner.

\subsection{Linear solutions for the case $r_m>0$}      \label{LSg0}

When $\gamma(r)$ is maximised off the disc axis, the eigenfunctions
are localised in its vicinity on the length scale $\Delta r\propto\lambda^s$,
while the corresponding eigenvalue decrement $\gamma_m-\Gamma$,
which measures the growth rate $\Gamma$, scales as $\lambda^{2s}$.
These two important qualitative measures of the non-local and local
eigensolutions are distinguished by $s=1/3$ and $s=1/2$, respectively (see
(\ref{scaling})  and Paper I). Though these scalings
are derived for our present  case $r_m>0$, they appear to hold also
for the case $r_m=0$ of Section \ref{LS} as Fig.~\ref{fig:lambda-var} illustrates.
Another important distinction of the non-local eigenfunctions is that they
decay only algebraically at $r\to\infty$ (Soward, 1992a,b), in contrast to the
exponential decay of the local eigenfunctions (e.g., Moss {\it et  al.}, 1998).

As an illustrative example we consider the local growth rate
\begin{equation}
   \gamma(r) =
   \left\{
      \barr{ll}
         \gamma_m r(2r_m-r) /r_m^2, & \quad r\leq  2r_m\,, \\
         \gamma_\infty,               & \quad r> 2r_m \,
      \earr
   \right.
\label{gamma1}
\end{equation}
similar to (\ref{gammaex1}) with $\gamma_m''=-2\gamma_m/r_m^2$ giving
$\epsilon=(\lambda r_m^2)^{1/3}$.
The leading eigenfunction for the parameter values
$\lambda=0.01$, $\gamma_m=1$, $\gamma_\infty=0$, $r_m=0.25$ and $R=4$
is shown in Fig.~\ref{fig:steady-lin}. Other values of $\gamma_\infty$ are employed
in Fig.~\ref{fig:steady-nonlin} below.

\begin{figure}
   \bcent
      \epsfig{figure=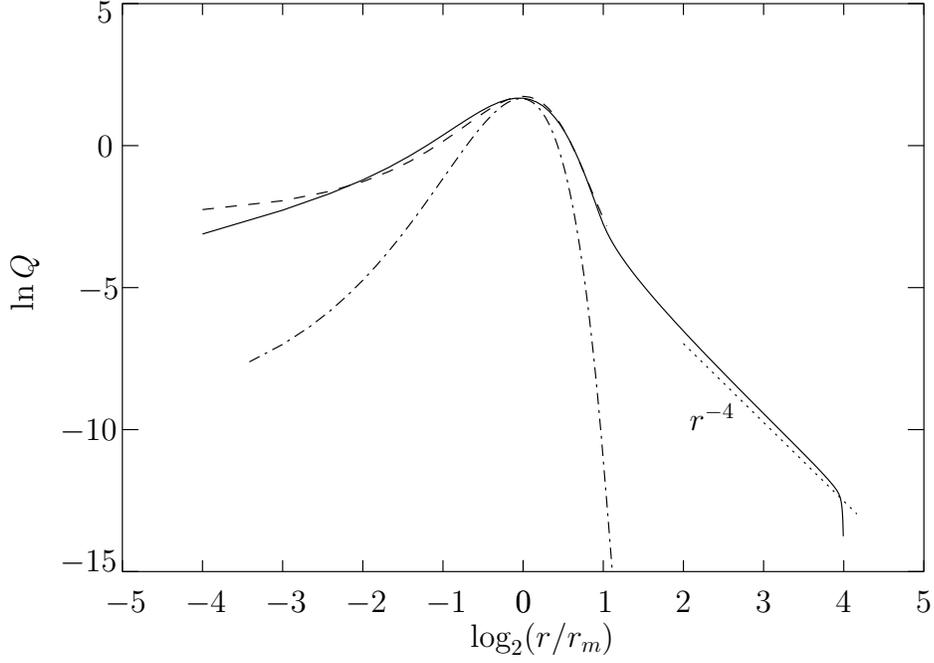, scale=0.85}
      \caption{ \label{fig:steady-lin}
The leading eigenfunctions $Q(r)$ appropriate to
$\gamma(r)$ defined by (\ref{gamma1}) with
the parameter values
$\lambda=0.01$, $\gamma_m=1$, $\gamma_\infty=0$, $r_m=0.25$ and $R=4$
for the `local' (\protect\ref{local}) (dash-dotted)
and the non-local (\protect\ref{nonlocal}) (solid) problems;
remember that $\gamma=\gamma_\infty$ on $1<\log_{2}(r/r_m)<4$.
The local solution has an exponential tail for large $|r-r_m|$, 
whereas the non-local solution exhibits an algebraic tail.
The dashed curve shows the WKB asymptotic solution
(\ref{intsol}) with the scaling (\ref{scaling}).
The dotted line shows the far field asymptote (\ref{largevarpi}) or
(\ref{gexex1}), valid for $r\gg 2r_m$ .
}
\ecent
\end{figure}

We compare our eigenfunction with the asymptotic solution
(\ref{intsol}) valid close to $r_m$ on the length scale
$\epsilon=(\lambda r_m^2)^{1/3}$, shown with dashed
curve in Fig.~\ref{fig:steady-lin}.
For $\epsilon\ll |r-r_m|\ll r_m$ it determines the
power law behaviour $Q\propto  |r-r_m|^{-4}$ (see (\ref{largevarpi}))
and this remains throughout the range $0\le r< 2r_m$
over which $\gamma(r)$ has a quadratic profile.
This actually links to a line dipole source, for which the poloidal magnetic
field decays as $|r-r_m|^{-2}$.
For $r>2r_m$, where $\gamma(r)=\gamma_\infty$ is a constant,
we have instead (\ref{gex}), which takes the form
\begin{equation}
q(r)\,\approx\,-\,\frac{\lambda {\cal Q}}{\gamma_m-\gamma_\infty}
\,r_m\frac{\partial^2 G}{\partial r \partial r_m}(r,r_m)\,,
\label{gexex}
\end{equation}
where $r_m\partial^2 G/\partial r \partial r_m(r,r_m)$ has the
asymptotic representation (\ref{diquadru}) giving
\begin{equation}
q(r)\,\approx\,\frac{3 \lambda r_m^2{\cal Q}}{2(\gamma_m-\gamma_\infty)r^4}
\label{gexex1}
\end{equation}
similar to (\ref{diquadru}). This is the reason for the $r^{-4}$
power law evident in Fig.~\ref{fig:steady-lin}
for $r>2r_m$ and corresponds
to the algebraic decay of its quadrupole source.
This asymptotics improves as $\gamma_\infty$
decreases, and the exponent $-4$
was generally found in the numerical
results independent of the value of $\gamma_\infty$,
at least for $\gamma_\infty\leq0$.

The main points to note are that the asymptotic solution
(\ref{intsol}) generated by the Airy function
approximates the numerical solution in the vicinity of $r_m$.
The nature of Fig.~\ref{fig:steady-lin} (namely, its relatively narrow relevant
range in $r$) makes it impossible
to easily identify the power law
$|r-r_m|^{-4}$, given by (\ref{largevarpi}),
inside the source region $0\le r<2r_m$.
Whether or not it is there, the power law
$r^{-4}$ outside the source region $r>2r_m$
has a different explanation for its similar power law exponent of
$-4$ given by (\ref{gexex1}),
specifically stemming from the quadrupole nature of the source. We also note that
the horizontal and vertical components of magnetic field scale differently with $r$
in the far field: 
$B\propto Q\propto r^{-4}$, 
$B_r=-\partial A/\partial z\propto r^{-4}$,
and $B_z=r^{-1}\partial(rA)/\partial r\propto dQ/dr\propto r^{-5}$. 
Thus, the
magnetic field becomes more horizontal with $r$ at large distances from the
dynamo active region.

      \subsection{Steady states}  \label{SS}

A simple heuristic form of nonlinearity in the radial dynamo equation
(\ref{local}), resulting from the standard from of $\alpha$-quenching,
$\alpha=\alpha_0\left(1+B^2/B_0^2\right)^{-1}$, with $\alpha_0$ the background
unquenched value, was derived by Poezd {\it et al.} (1993) who suggest
that $\gamma(r)$ is to be replaced by
\begin{equation}        \label{quench}
   \gamma^{\rm(n)}(r,t)  =  \gamma(r)
   \left[ 1-Q^2(r,t)/Q_0^2(r) \right],
\end{equation}
where $Q_0(r)$ is essentially equal to $B_0$, but with the
possible $z$-dependence of $B_0$ averaged out.
In this section, we compare the local and non-local steady states
using this form
with $Q_0(r)=1$ in both (\ref{local}) and (\ref{nonlocal}),
together with $\gamma(r)$ of the form (\ref{gamma1})
where $\gamma_\infty$ was varied to explore properties of the solution
at $r\ga r_1$.

Taking the eigenfunctions of Sect.~\ref{LSg0} as initial conditions, we solved
the Cauchy problem for both the local and non-local dynamo equations using the
Runge--Kutta method. This was performed also for the corresponding linear
equation in order to verify the accuracy of the computations.

\begin{figure}
   \bcent
         \epsfig{figure=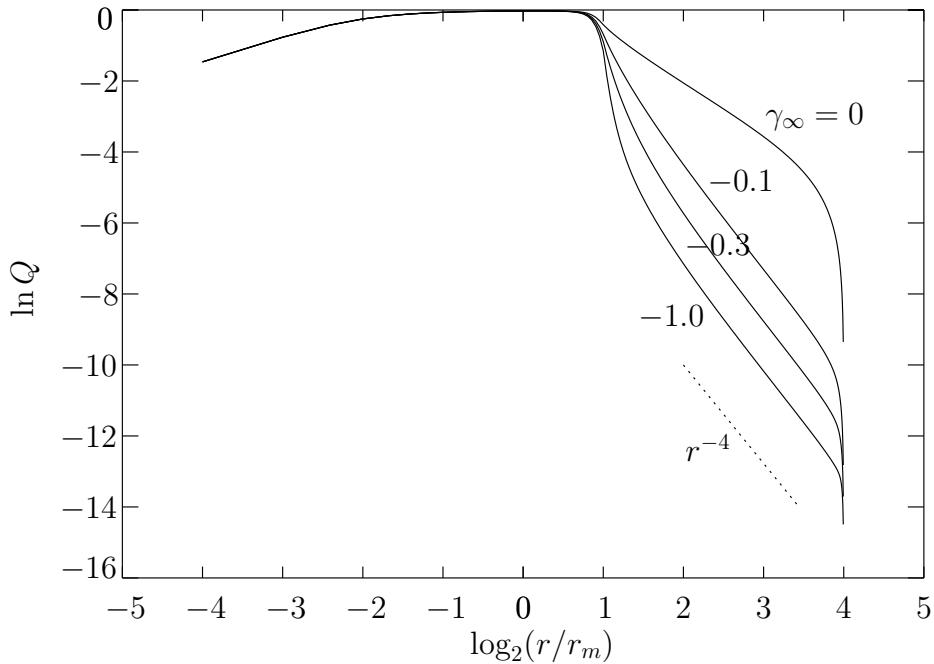, scale=0.85}
      \caption{ \label{fig:steady-nonlin}
Steady-state nonlinear non-local solutions obtained for
$\gamma(r)$ of the form (\protect\ref{gamma1}) with
$R=4$, $r_m=0.25$, $\lambda=0.01$, and $\gamma_\infty=0,\,-0.1,\,-0.3,\,-1.0$
(with the value of $\gamma_\infty$ shown near the corresponding curve).
The algebraic tail remains similar to that in the linear
solutions, $Q\sim r^{-4}$, with the exponent independent
of $\gamma_\infty$, although the magnitude of $Q$
in the tail reduces as $\gamma_\infty$ decreases.
}
   \ecent
\end{figure}

Figure \ref{fig:steady-nonlin} shows steady states for the non-linear regime.
The non-linear solution clearly
has a broader distribution in $r$ than the associated eigenfunction
in Fig.~\ref{fig:steady-lin}b.
The algebraic tail $Q\sim r^{-4}$ characteristic of the non-local
eigenmodes persists in the nonlinear solutions. Its functional form is
independent of the local value of $\gamma$, as shown by our
solutions with various values of $\gamma_\infty$.
This is consistent with the fact that the algebraic tail results
from the non-local magnetic connection with the dynamo-active region at smaller
$r$. However, the
magnitude of $Q(r)$ at large $r$ is affected by the value of $\gamma_\infty$.
This effect can be estimated from (\ref{gex}) with $\Gamma=0$, which
yields $q(r)\propto {\mathcal Q}/\gamma_\infty$.  The values of $q$ over the
appropriate range of $r$ in Fig.~(\ref{fig:steady-nonlin}) agree with this
to within 30\%
for $\gamma_\infty=-0.3$, and 20\%
for $\gamma_\infty=-1$.

      \subsection{Propagating Magnetic Fronts}  \label{PMF}
If the local growth rate, $\gamma_0(r)$, is localized in a
limited radial domain, growing magnetic field distribution
of the local kinematic dynamo problem, (\ref{local}), propagates at a finite
velocity, $V$, into the dynamo-inactive region.  This phenomenon
was discussed by Moss {\it et al.} (1998, 2000), Petrov {\it et al.}
(2001), Petrov (2002) and Fedotov {\it et al.} (2003).
The eigenfunctions of (\ref{local}) decay
exponentially at large $r$,
$Q(r)\propto\exp(\Gamma t-r/r_0)$, where $r_0=(\beta/\Gamma)^{1/2}$
in dimensional variables. The position $r_{\rm f}$
where $Q=\mbox{const}$ changes with time as $r_{\rm f}=r_0 \Gamma t
=(\beta\Gamma)^{1/2}t$,
i.e., the magnetic front propagates at a constant speed
$V=(\Gamma\beta)^{1/2}$ (Moss {\it et al.,} 1998). The front propagation
is strongly affected by the asymptotic behaviour of $Q(r)$ at $r\gg1$:
for the non-local solution,
we have
$Q\propto (r/r_0)^{-4}\exp{(\Gamma t)}$, so that
$Q=\mbox{const}$ now yields $r_{\rm f}=r_0\exp{\Gamma t/4}$, i.e.,
the magnetic front
propagation is exponentially fast, with a speed increasing
with time
as $V\propto\exp{\Gamma t/4}$. This is a direct consequence of the non-local
coupling
of the inner regions, where the dynamo is strong, with the outer
regions where the front propagates: since magnetic fields propagates
at infinite speed through vacuum, magnetic field far from the
dynamo region responds instantaneously to the exponential growth of magnetic
field in the inner parts of the disc.

\begin{figure}
   \bcent
      \epsfig{figure=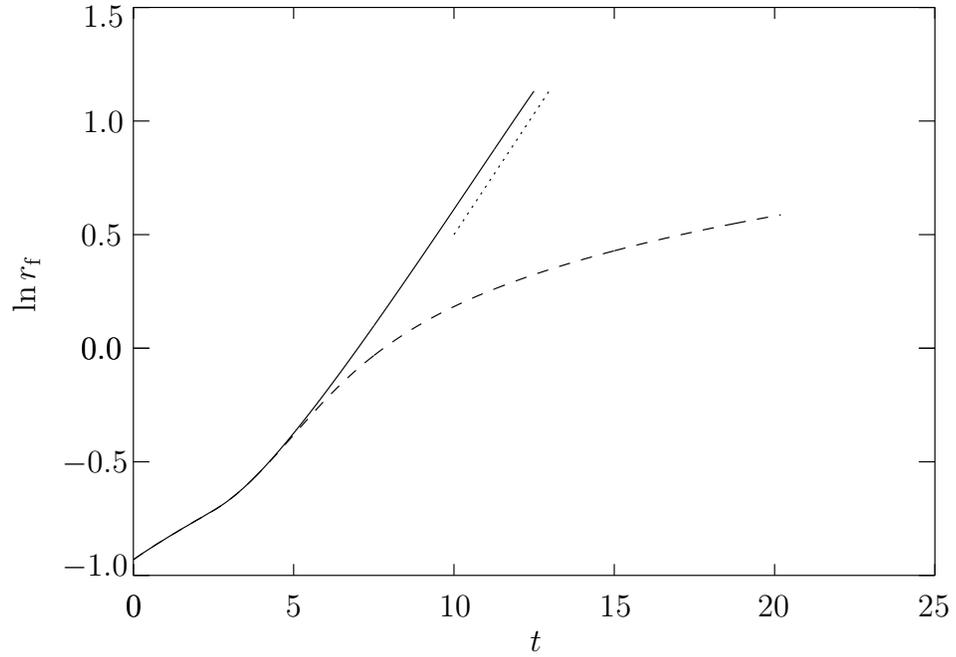, scale=0.85}
      \caption{ \label{fig:ext-fronts}
         The position of a propagating magnetic front versus time
for the linear (solid) and nonlinear (dashed) solutions of (\protect\ref{nonlocal})
with $\gamma(r)$ of the form (\protect\ref{gamma1}) with
the same parameters as in Fig.~\ref{fig:steady-lin}b and $\Gamma=0.852$.
The dotted line represents $\exp(\Gamma t/4)$.
      }
   \ecent
\end{figure}

These heuristic arguments are confirmed by numerical solutions illustrated
in Fig.~\ref{fig:ext-fronts} where we show the position of the magnetic front
defined here as a position $r=r_\mathrm{f}$ where $Q(r_\mathrm{f})=10^{-3}$.
The eigenfunction of the nonlocal problem was used as the initial condition,
normalised such that its maximum was $10^{-2}$.
The eigenfunction
decreases with $r$ relatively quickly for $r_m/2<r<2r_m$ (within
the dynamo region) and then slower at $r>r_m$ (where $\gamma(r)=\gamma_\infty\leq 0$)
--- see Fig~\ref{fig:steady-lin}.
Correspondingly, the front first propagates slowly
through the region where the solution
rapidly decreases away from its maximum, but
the front speeds up as it reaches the region of the algebraic tail.
Front propagation is inhibited by nonlinear effects
as they suppress the exponential growth of the solution and then establish a
steady state.

   \section{Conclusions and Discussion}

Arguably the most important property of the non-local
solutions discussed above is the slow, algebraic decay of both
kinematic and steady-state mean magnetic fields at large
distances, along radius, from the field maximum. This feature can be
important for the understanding of large-scale magnetic fields
observed at large distances from the centres of spiral galaxies where
both low gas density and reduced intensity of interstellar turbulence
would preclude any strong magnetic fields. However, Han {\it
et al.} (1998) have found indications, albeit not very strong, that
the large-scale magnetic field in the disc of the Andromeda nebula
(M31) at distances as large as 25\,kpc
is almost as strong as at a radius of 10\,kpc. Beck (2003) argues that
magnetic field
in the galaxy NGC~6946 (one of the better studied nearby spiral galaxies)
decreases with galactocentric radius unexpectedly slowly, so it can
become dynamically important in the outer Galaxy. If approximated by an exponential
profile, the energy density of the total (regular plus random) magnetic field has the radial
length scale of perhaps as large as 8\kpc. Meanwhile, the gas density
decreases exponentially at a length scale of 2--3\,kpc; if the turbulent velocity
does not vary much with radius, this length scale applies to the turbulent
energy density as well. Thus, the observations indicate that the large-scale
magnetic field may decay with $r$ slower than the turbulent energy density.

The algebraic tail produced by the non-local magnetic effects can be
responsible for the slow decrease of the observed magnetic field with radius.
It would be important to test this idea with numerical solutions for a
disc dynamo embedded in a poorly conducting halo.

The assumption that the disc is surrounded by vacuum is
essential for the nature of the long-distance behaviour of the
dynamo-generated magnetic field. Real galactic discs are surrounded
by turbulent halos, and the vacuum boundary conditions are
justified if the time scale of the (turbulent) magnetic field
diffusion above the disc is significantly shorter that through the
disc, i.e., if
\[
\frac{\tau_\mathrm{h}}{\tau_\mathrm{d}}
=\frac{\beta_\mathrm{d}}{\beta_\mathrm{h}}\,\frac{L^2}{R^2}\ll1\;,
\]
where $\beta_\mathrm{h}$ and $\beta_\mathrm{d}$ are the
turbulent magnetic diffusivities in the halo and the disc,
respectively, $R$ is the radial length scale of the magnetic field in
the disc, and $L$ is the length of a magnetic line that leaves the
disc through its surface and then returns to the disc with the
radial distance $R$ between its footpoints. As argued by Poezd
{\it et al.} (1993), $\beta_\mathrm{h}/\beta_\mathrm{d}\simeq10$--30,
assuming that both the turbulent velocity and scale are in the halo
are 3--5 times larger than in the disc.
With $L/R$ of order unity, this yields $\tau_\mathrm{h}/\tau_\mathrm{d}\ll1$;
for $L/R\simeq2$ we obtain
$\tau_\mathrm{h}/\tau_\mathrm{d}\simeq0.1$--$0.5$, a value arguably
small enough to believe that the magnetic coupling through
the halo is important.

Another condition for the algebraic tail to be established
is that the magnetic propagation time through the halo is short enough
in comparison with the galactic lifetime,
$\tau_\mathrm{h}\la10^{10}\yr$. Using for $L$ the radial
length scale of a kinematic non-local magnetic mode in the disc,
$L\simeq\lambda^{-2/3}h$,
i.e., assuming that the length of magnetic lines joining two positions
in the disc through the halo does not differ strongly from the radial separation
of the positions), where $h\simeq500\p$ is the scale
height of the galactic ionized disc and $\lambda=0.05$--0.1 is its aspect ratio, we obtain
$\tau_\mathrm{h}\simeq(5$--$40)\times10^8\yr$, with
$\beta_\mathrm{h}=(10$--$30)\beta_\mathrm{d}$ and
$\beta_\mathrm{d}\simeq10^{26}\cm^2\s^{-1}$.
Considering the uncertainty of the halo parameters, this estimate seems to
indicate that the algebraic tail can be established in
spiral galaxies, especially in those with strong halo turbulence or small
disc.

\section*{Acknowledgements}
We are grateful to W.~Dobler for useful discussions. This work was
supported by the Royal Society, the NATO collaborative research grant
CRG1530959, and the PPARC Grant PPA/G/S/2000/00528.

\section*{References}
\begin{description}

\item Abramowitz, M.\ and Stegun, I.~A.\ (Eds), {\it Handbook of Mathematical
Functions,} Dover Publ., New York (1965)

\item Beck, R., ``The Role of Magnetic Fields in Spiral Galaxies,'' in: {\it
From Observations to Self-Consistent Modelling of the ISM in Galaxies} (Eds
M.\ Avillez \& D.\ Breitschwerdt), Kluwer, Astrophys.\ \& Space Sci.\ Ser., in
press (astro-ph/0212288) (2003).

\item Bullard, G. E. and Gellman, H.\
``Homogeneous dynamos and terrestrial magnetism,''
{\it Phil. Trans. R. Soc. Lond.} A, {\bf 250}, 543-585 (1954).

\item Fedotov, S., Ivanov, A.\ and Zubarev, A., ``Non-local mean-field
dynamo theory and magnetic fronts in galaxies,''  {\it Geophys.\ Astrophys.\
Fluid Dynam.}, {\bf 97}, 135-148 (2003).

\item Gradshteyn, I.~S.\ and Ryzhik, I.~M.\  {\it Table of Integrals, Series and
Products,} Academic Press (2000)

\item Han, J.~L., Beck, R.\ and Berkhuijsen E.~M., ``New clues to the magnetic
field structure of M31,'' {\it Astron.\ Astrophys.,} {\bf335}, 1117--1123
(1998).

\item Krause, F.\ and R\"adler, K.-H., {\it Mean-Field Magnetohydrodynamics
and Dynamo Theory,} Pergamon, Oxford (1980).

\item Moffatt, H.~K., {\it Magnetic Field Generation in Electrically
Conducting Fluids,} Cambridge Univ.\ Press (1978).

\item Moss, D., Petrov, A.\ and Sokoloff, D., ``The motion of
magnetic fronts in spiral galaxies,'' {\it Geophys.\ Astrophys.\
Fluid Dynam.} {\bf 92}, 129--149 (2000).

\item Moss, D.\ and Shukurov, A., ``Accretion disc dynamos opened up by
external magnetic fields,'' {\it Astron.\ Astrophys.,} 
{\bf 413}, 403-414 (2004).

\item Moss, D., Shukurov, A.\ and Sokoloff, D., ``Boundary effects
      and propagating magnetic fronts in disc dynamos,'' {\it
Geophys.\ Astrophys.\ Fluid Dynam.} {\bf 89}, 285--308 (1998).

\item Petrov, A.~P., ``Travelling contrast structures in various hydromagnetic
dynamo models,'' {\it Matematicheskoe Modelirovanie} {\bf 14}, 95--104 (2002) (in Russian).

\item Petrov, A.~P., Sokoloff, D.~D.\ and Moss, D., ``Magnetic fronts
and two-component asymptotics for galactic magnetic fields,''
{\it Astron.\ Zh.} {\bf78}, 579--584 (2001).
English translation: {\it Astron. Rep.}, {\bf 45}, 497--501 (2001).

\item
Poezd, A., Shukurov, A.\  and Sokoloff, D., ``Global magnetic
         patterns in the Milky Way and the Andromeda nebula,'' {\it
Mon.\ Not.\ Roy.\ Astron.\ Soc.} {\bf 264}, 285--297 (1993).

\item
Press, W.~H., Flannery, B.~P., Teukolsky, S.~A.\ and Vetterling, W.~T.,
{\it Numerical Recipies}, Sect.\ 18.3.2, Cambridge Univ.\ Press, Cambridge (1993).

\item Priklonsky, V., Shukurov, A., Sokoloff, D.\ and Soward, A.,
``Non-local effects in the mean-field disc dynamo. I.  An asymptotic
 expansion,'' {\it Geophys.\ Astrophys.\ Fluid Dynam.} {\bf 93}, 97--114
(2000). (Paper I.)

\item R\"adler, K.-H.\ and Wiedemann, E. 
``Mean field models of galactic dynamos admitting axisymmetric and 
non-axisymmetric magnetic field structures,'' 
in: {\it Galactic and intergalactic magnetic fields}
(Eds R. Beck, P. P. Kronberg \& R. Wielebinski), Kluwer,
107-112 (1990). 

\item Ruzmaikin, A.~A., Shukurov, A.~M.\ and Sokoloff, D.~D., {\it
 Magnetic Fields of Galaxies}, Kluwer, Dordrecht (1988).

\item
Ruzmaikin, A.~A., Sokoloff, D.~D.\ and Shukurov, A.~M., ``Magnetic
field distribution in spiral galaxies,'' {\it Astron.\ Astrophys.}
{\bf148}, 335--343  (1985).

\item Soward, A.M., ``A thin disc model of the Galactic dynamo,''
   {\it Astron.\ Nachr.} {\bf 299}, 25--33 (1978).

\item Soward, A.M., ``Thin aspect ratio $\alpha\Omega$-dynamos
   in galactic discs and stellar shells,'' in: {\it Advances in Nonlinear Dynamos}
(Eds A.~Ferriz-Mas \& M.~N\'u\~nez), Taylor \& Francis, London, pp.\ 224--268 (2003).

\item Soward, A.M., ``Thin disc $\alpha\omega$-dynamo models.
   I.  Long length scale modes,''  {\it Geophys.\ Astrophys.\ Fluid
   Dynam.} {\bf 64}, 163--199 (1992a).

\item Soward, A.M., ``Thin disc $\alpha\omega$-dynamo models II.
   Short length scale modes,''  {\it Geophys.\ Astrophys.\ Fluid Dynam.}
   {\bf 64}, 201--225 (1992b).

\item Zeldovich, Ya.~B., Ruzmaikin, A.~A.\ and Sokoloff, D.~D., {\it
The Almighty Chance,} World Scientific Publ., Singapore (1990).

\end{description}

\section*{Appendix A. The evaluation of the elliptic integral
(\ref{Kdef}) and its derivative}
\setcounter{equation}{0}
\renewcommand\theequation{A\arabic{equation}}
To evaluate the elliptic integral (\ref{Kdef}),
where we have introduced $\kappa$ via $m=\sin^2\kappa$,
and used an iterative method described by Abramowitz and Stegun (1965):
for the staring values
\begin{equation}                        \label{start}
a_0=1\;,\qquad b_0=\cos\kappa=\sqrt{1-m}\;,
\end{equation}
we calculate
\begin{equation}                \label{iter}
a_{n+1}=\sfrac{1}{2}(a_n+b_n)\;,\qquad b_{n+1}=\sqrt{a_nb_n}
\end{equation}
to obtain
\begin{equation}                \label{res}
K(m)=\frac{\pi}{2a_\infty}\;.
\end{equation}

The derivative $dK/dm$ is calculated using a generalization of this
algorithm suggested by W.~Dobler (unpublished). We perturb the system
above, $m\to m+\delta m,\ a_n\to a_n-\kappa_n\delta m,\
b_n\to b_n-\zeta_n\delta m,$ where $\delta m\ll1$. The starting values
$\kappa_0$ and $\zeta_0$ are obtained from the Taylor expansions of
(\ref{start}):
\[
\kappa_0=0\;,\qquad \zeta_0=\frac{1}{2\sqrt{1-m}}\;,
\]
and similarly from (\ref{iter}):
\[
\kappa_{n+1}=\sfrac{1}{2}(\kappa_n+\zeta_n)\;,\qquad
\zeta_{n+1}=\frac{a_n\zeta_n+\kappa_n b_n}{2b_{n+1}}\;.
\]
Perturbing (\ref{res}) with respect to $m$, we obtain
\[
\frac{dK}{dm}=\frac{\pi}{2a_\infty^2}\kappa_\infty\;.
\]
Typically, one or two iterations are sufficient to
calculate $K(m)$, $dK/dm$ to the eighth decimal place.

\section*{Appendix B. The regularization of the singular integral kernel
$W(r,r')$, (\protect\ref{GreenJ})}\label{sect:regularisation}
\setcounter{equation}{0}
\renewcommand\theequation{B\arabic{equation}}
Equation (\ref{nonlocal}) is solved numerically for
$0\leq r\leq R$, with suitably large $R$. The integral kernel of this
equation, $W(r,r')$ has a singularity along the
line $r=r'$.  Simple quadrature methods may display poor
convergence if such singularities are ignored and do not apply
at $r=r'$. These difficulties are resolved using
the subtraction of the singularity (Press {\it et al.,} 1993)
by performing the following transformation:
\[
   [\Gamma - \gamma(r)] q(r)\,=\,\frac{\lambda \eta(r)}{r}
   \int_0^R W(r,r') \widehat{L}_{r'}(q(r'))\,dr'
\qquad\qquad\qquad\qquad
\]
\beq
\qquad
    = \,\frac{\lambda \eta(r)}{r}
   \left\{
   \int_0^R W(r,r')
      \left[
         \widehat{L}_{r'}(q(r')) - \widehat{L}_r(q(r))
      \right]dr'\,
      + \widehat{L}_r(q(r)) M(r)                        \nonumber
   \right\},
\label{eq:sub_of_sing}
\eeq
where
\[
\widehat{L}_r(q(r))\,\equiv\,\pder{}{r}{}\left[\frac{1}{r}\pder{}{r}{}\Bigl(rq(r)\Bigr)\right],
\]
\beq
   M(r)\,\equiv\,\int_0^R W(r,r')\,dr'\,
   = \,r \int_0^R r' \left[\int_0^\infty J_1(kr) J_1(kr')\,dk\right]dr'\,.
\label{eq:M_r}
\eeq
As a result, the singularity in the integral operator at $r=r'$ is
regularized since $\widehat{L}_{r'}(q(r'))-\widehat{L}_r(q(r))=0$ at $r=r'$.
The additional term
$\lambda \eta(r)M(r)r^{-1}\widehat{L}_r(q(r))$
describes a modified radial diffusion with diffusivity
$\lambda \eta(r)M(r)r^{-1}$.

For finite, non-vanishing values of $r$, $M(r)$ can be calculated as
follows. Making the substitution $\sigma = kr$, we obtain
\[
   M(r)\, =\,\int_0^R r' \left[\int_0^\infty J_1(\sigma) J_1(\sigma r'/r)\,d\sigma\right]dr'
\]
and with $x =r'/r$ we find
\[
   M(r)\, = \,r^2\int_0^{R/r} x \left[\int_0^\infty J_1(\sigma) J_1(\sigma x)\,d\sigma\right]dx
       \,=\, r^2 \widetilde{M}(r)\,,
\]
where
\[
  \widetilde{M}(r)\,= \,\int_0^{R/r} W(1,x)\,dx
\]
is a universal function of $r$ that need only be calculated once.
For a mesh point $r_n$, $n\leq N$, $r_N=R$,  we
have
\begin{eqnarray}
   \nonumber
   \widetilde{M}_N &=& \int_0^1 W(1,x)\,dx\,,\\
   \widetilde{M}_{n-1} &=& \widetilde{M}_n + \int_{R/r_n}^{R/r_{n-1}}
   W(1,x)\,dx\,,
   \qquad n=N,\dots 2\;,
   \label{recurr}
\end{eqnarray}
where $\widetilde{M}_n = \widetilde{M}(r_n)$.
Using an (open) quadrature method, we find
$\widetilde{M}_N\approx0.31409623$, the remaining
$\widetilde{M}_n$, $n<N$ depend
on the descretisation.  From the recurrence relation (\ref{recurr}),
$\widetilde{M}_n$, and hence $M_n$, may be
precomputed by a single integration over the range $1$ to
$R/r_1$.

We now take centered differences on the regularized equation
(\ref{eq:sub_of_sing}).
The point $j=i$ in the corresponding sum is omitted since there
$[\widehat{L}_{r_j}(q(r_j))-\widehat{L}_{r_i}(q(r_i))] = 0$.
This leads to the eigensystem
$\Gamma \vec{Q} = A \vec{Q}$
which we solve using the QR algorithm.

\end{document}